\documentclass[10pt]{iopart}

\usepackage{graphicx}%
\usepackage{dcolumn}%
\usepackage{bm}
\usepackage{color}
\date{\today}
\RequirePackage{lineno}
\begin{document}
\title{Applications and non-idealities of submicron Al-AlOx-Nb tunnel junctions}
\author{J. K. Julin and I. J. Maasilta}
 
 \address{Nanoscience Center, Department of Physics, University of Jyvaskyla, P. O. Box 35, FIN-40014 Jyvaskyla, Finland}
\ead{maasilta@jyu.fi}

\begin{abstract}
We have developed a technique to fabricate sub-micron, $ 0.6 \mu$m$ \times 0.6 \mu$m Al-AlOx-Nb tunnel junctions using a standard e-beam resist, angle evaporation and double oxidation of the tunneling barrier, resulting in high quality niobium, as determined by the the high measured values of the critical temperature $T_C \sim 7.5$ K and the gap $\Delta \sim 1.3 $ meV. The devices show great promise for local nanoscale thermometry in the temperature range 1 - 7.5 K. Electrical characterization of the junctions was performed at sub-Kelvin temperatures both with and without an external magnetic field, which was used to suppress superconductivity in Al and thus bring the junction into a normal-metal-insulator-superconductor (NIS) configuration. We observed excess sub-gap current,  which could not be explained by the standard tunneling theory. Evidence points towards materials science issues of the barrier or Nb/AlOx interface as the culprit. 

\end{abstract}

\submitto{\SUST}
\maketitle
\ioptwocol

\section{Introduction}

Superconducting tunnel junctions have proven to be valuable devices widely used in quantum bits, \cite{squbit2} sensitive magnetometers \cite{Clarke}, metrological applications, \cite{RevModPhys.85.1421}, sensitive thermometers \cite{SINIS} and radiation detectors\cite{enss}, and even electronic coolers \cite{muhonen}. 
Most devices so far have used either aluminum or niobium as the superconductor, due to the ease of fabrication and extensive know-how of their processing. Aluminum is particularly widely used, as its thermal oxide is of high quality, and thus simple room temperature oxidation of an Al film will produce a high quality tunnel barrier. For some applications, however, the critical temperature $T_C$ and the superconducting gap $\Delta$ of Al are too low (typically $\sim 1.5$ K and $\sim 0.2$ meV, respectively) so that niobium is desired. However, it does not have the luxury of a good thermal oxide, so most Nb superconductor-insulator-superconductor (SIS) junctions are fabricated with a trilayer process \cite{gurvitch}, where a thin ($\sim 10$ nm) Al layer must be deposited for the barrier formation.   

The trilayer process for Nb junctions is well suited for devices where the junction size is above the sub-micron length scale. For smaller devices, it becomes easier to use the angle evaporation technique \cite{dolan}, which avoids multilayer lithography, deposition of insulators, and possible damage problems during the plasma etching  in the trilayer process. A downside of the angle evaporation is that an e-beam sensitive polymer resist mask is used. This does not cause problems for Al based junctions, but is not as staightforward for Nb, as the quality of evaporated Nb is quite sensitive to the chamber and substrate conditions, such as vacuum level, evaporation speed, substrate to Nb crucible distance, and especially the type of resist used \cite{Kim,Pekola1999653,Dolata,Harada,Dubos,Hoss,Im}. Problems with standard polymer resists such as poly-methyl-methacrylate (PMMA) are typically attributed to the decomposition and outgassing of the resist during Nb evaporation, leading to suppression of $T_C$ and $\Delta$ \cite{Kim,Pekola1999653,Dolata,Harada,Im} from the bulk values  $T_C \sim 9.2$ K and  $\Delta \sim 1.5$ meV.   

Here, we have successfully fabricated sub-micron Al/AlOx/Nb tunnel junctions using angle evaporation and standard PMMA/P(MMA-MAA) double layer resists, obtaining a high $T_C \sim 8$ K and a high $\Delta \sim 1.3$ meV. Typically, such high quality sub-micron Nb tunnel junctions are fabricated with a special resist capable of withstanding higher thermal loads \cite{Dolata02,pashkin}, and only one other report of such high-quality Al/AlOx/Nb junction fabrication with PMMA exists \cite{paraoanu}, with device applications as superconducting single-electron transistors \cite{toppari} in the ultra-small junction size ($0.1 \times 0.1 \mu$m$^2$) scale. In fact, Refs. \cite{paraoanu,toppari} used the same exact evaporator as here. In contrast to Refs. \cite{paraoanu,toppari}, we use a different geometry allowing also a combination of smaller and larger junction sizes to be fabricated, and mostly focus on the junction properties in the normal-metal-insulator-superconductor (NIS) state, by either using temperatures above the Al $T_C \sim$ 1.3 K, or by applying an appropriate external magnetic field ($B$) to suppress the superconductivity in Al, but not in Nb. The interest in the NIS geometry stems from thermometry \cite{SINIS}, cooling \cite{muhonen}, and metrological applications \cite{RevModPhys.85.1421}.

 We find that although the Nb quality is high right at the junction as judged by the high $\Delta$, we observe excess sub-gap current in the NIS state which cannot be explained by the simplest theories of broadening of the superconducting density of states \cite{PhysRevLett.53.2437,PhysRevLett.105.026803,mitrovic}. This result is in agreement with previous studies in Al/AlOx/Nb \cite{toppari} and in Al/AlOx/V \cite{shimada} junctions, but in contrast with our previous work with Nb-Al-AlOx-Cu \cite{minna} or other similarly fabricated NIS junctions of NbN \cite{nbn} or TaN \cite{tan}, where the simple Dynes broadening \cite{PhysRevLett.53.2437,PhysRevLett.105.026803} {\em does} explain the data. We comment at the end on the possible explanations for this difference.

\section{Sample fabrication}

If the goal is to fabricate a Nb-based NIS junction, there are two possibilities with angle evaporation: either the Nb is evaporated first or last. Let's discuss the pros and cons of the two options. If Nb is evaporated first, the high evaporation temperature will not damage other films and especially the tunneling barrier. However, Nb can form several oxides \cite{Halbritter1987}, out of which only one, Nb$_2$O$_5$, is a good insulator, so that thermal oxidation of Nb produces bad quality junctions. That's why typically one has to evaporate a thin Al layer on top of the Nb, and oxidize the Al instead of the Nb \cite{gurvitch}. For tri-layers, this works well, but with angle evaporation an added difficulty is that the Al layer must cover the Nb electrode  well everywhere to prevent a direct leakage path from the Nb layer. 

In our previous work \cite{minna}, this coverage problem was solved with a special geometric electrode design.
However, this approach leads to issues with the proximity effect of the Al layer, which degrades the superconducting properties of the Nb, by reducing and possibly also broadening the gap. It was also difficult to prevent the formation of an Al gap feature in the data,  which was believed to be caused by parallel tunneling paths from purely Al areas of the tunnel junction \cite{minna}. The highest gap we  observed in Nb-Al-AlOx-Cu junctions fabricated this way was $\Delta \sim 1$ meV, with a Dynes broadening \cite{PhysRevLett.53.2437,PhysRevLett.105.026803} parameter $\Gamma/\Delta \sim 5 \times 10^{-2}$. 

The second option is to evaporate the normal metal electrode first. Typically, the same barrier formation problem then exists, unless the normal metal is Al or its alloy \cite{Ruggiero2004}. The proximity effect problem is solved, but the question remains, does the hot, evaporated Nb cause damage or degradation of the underlying tunneling barrier? We decided to study this second option here by fabricating Al-AlOx-Nb junctions, as some preliminary success was reported before \cite{paraoanu}.

\begin{figure}[h]
\centering
\includegraphics[width=0.9\columnwidth]{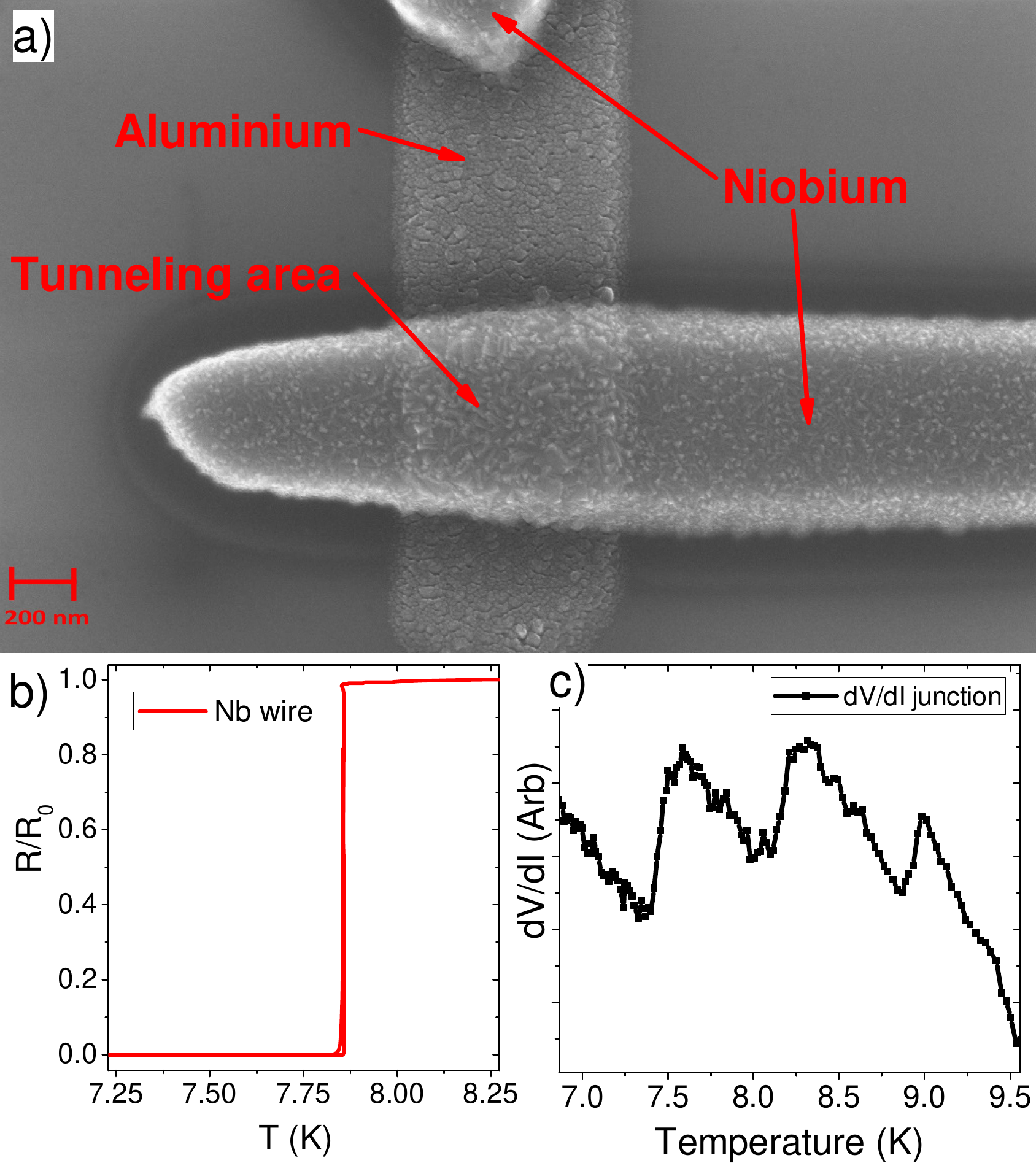}
\caption{(a) A scanning electron microscope (SEM) image of an Al-AlOx-Nb tunnel junction. (b) A measured transition of a 600 nm wide Nb wire. (c) The measured resistance of a Al-AlOx-Nb tunnel junction around the Nb transition temperatures, showing three separate drops at 7.5 K, 8.2 K and 9.0 K, due to three different width regions of the sample.}
\label{Al-Nb-SEM}
\end{figure}

The sample fabrication was done by electron beam lithography, by exposing a dual resist layer [PMMA and copolymer P(MMA-MAA)] on a surface of an oxidized silicon chip. This bi-layer forms the desired undercut structure, allowing angle evaporation. The junction geometry was such that the barrier is formed between the two metal layers crossing at 90 degree angle [Fig. 1(a)]. This way, different size junctions can be formed on the same normal metal island, facilitating different functionalities. In particular, for NIS thermometry and bolometry, one typically wants small area junctions to minimize self-heating, whereas larger area junctions operate much better as NIS coolers \cite{muhonen}. This is in contrast with hybrid NIS or fully superconducting single electron transistors,  where for all junctions the size is minimized to increase charging energy \cite{RevModPhys.85.1421,toppari,pashkin}. To simplify matters here, we concentrate only on the properties of single junctions. The linewidth for both electrodes was measured to be about 600 nm, as can be seen from the SEM image in Fig. \ref{Al-Nb-SEM} (a).

Prior to any metal deposition, the chip was cleaned in a reactive ion etcher with an O$_2$ plasma at 30-40 W power for 30 seconds,   with a pressure of 40 mtorr and 50 sccm flow rate, to reduce the effect of PMMA resist contamination on the exposed areas. All evaporations were done in an electron-beam evaporator with a chamber in ultra high vacuum ($P \sim 10^{-9}$ mbar), the Al oxidation was done in pure oxygen at room temperature in a separate loading chamber. 

The first aluminium layer was deposited to a thickness 45 nm from a large 65 degree angle, and then oxidized for 5 minutes in 1 bar pressure of pure oxygen. After that, we  tried depositing 80 nm of niobium on top at zero degrees from normal, with an evaporation speed of 0.5 nm/s. Note that this thickness is a factor of two more than in previous work on SET Al/AlOx/Nb junctions \cite{paraoanu,toppari, pashkin} and a factor of four more than in our own Cu-Nb NIS work \cite{minna}. This is important especially if we consider cooling applications, as thicker superconducting leads suffer less from non-equilibrium quasiparticle poisoning and back-tunneling \cite{clark2004,muhonen}, both of which deteriorate the cooler performance. However, in all trials, this standard technique led to a short between the electrodes, indicating that the hot Nb caused significant damage to the tunneling barrier. This failure, as compared to previous work, may have resulted in our larger junction sizes and thicker films, causing higher thermal load on the junction.

The solution to this problem was to use a second, $\sim 2$ nm aluminum layer, deposited after the oxidation of the first layer, and also thermally oxidize it for 20 minutes in 1 bar, to form a double oxide barrier \cite{houwman,tommy}. After such a step, the evaporated Nb did not short the barriers anymore, leading to functioning Al/AlOx/Nb tunnel junctions. 

\section{Electrical characterization}

To check the quality of the Nb, we also fabricated simple wires of the same width (600 nm) as the junction samples, and measured the superconducting transition, shown in Fig. \ref{Al-Nb-SEM} (b). A relatively high $T_C$ of 7.9 K is seen, and sometimes with similar conditions and widths $T_C$  as high as 8.6 K was observed. This is in stark contrast to narrow Nb wires made by several other groups using PMMA resists, where such width typically lead to $T_C$ suppression down to $\sim 2$ K \cite{Hoss,Dolata,Im}, but in agreement with earlier work using the same evaporator \cite{Kim}. We conclude that although clearly PMMA outgassing is detrimental for Nb, the use of PMMA is still possible, with the strength of the outgassing effect depending on the particular evaporator conditions, such as the distance between source and sample, for example.



The critical temperature of niobium used in the actual tunnel junction samples cannot be detected by searching for a sudden resistance drop to zero, since a high $\approx$50 kOhm tunneling resistance dominates over a much smaller wire resistance, thus making it more challenging to detect the small resistance change due to a superconducting transition. However, when accurately examined, it is possible to see three sharp and clear drops in  the differential resistance measurement plot, superimposed over a general increase of the tunneling resistance with lowering temperature, see Fig.\ref{Al-Nb-SEM} (c).These drops represent three  different parts of the niobium leads having different widths, and thus slightly different critical temperatures.  This is the most accurate data to measure the transition temperature of the niobium right at the tunnel junction region, which in the case of Fig. \ref{Al-Nb-SEM} (c) was 7.5 Kelvin (the smallest temperature corresponds the narrowest width structures in the sample).


For the rest of the measurements, a compact 3He-4He dilution refrigerator was used to measure the current-voltage characteristics of a single Al-AlOx-Nb tunnel junction at temperatures between 90 mK and 6.5 K. The sample was measured in a four probe configuration, with  Thermocoax cables and copper powder filters providing microwave filtering in the cryostat lines, to prevent overheating by noise and excessive photon-assisted tunneling.  The I-V measurements were performed with Ithaco 1201 voltage and  1211 current preamplifiers, and the voltage/current source was battery powered to reduce interference and noise. The refrigerator had a superconducting coil in the 4.2 K vacuum jacket, which was used to create a magnetic field  perpendicular to the sample substrate plane, and thus perpendicular to the tunnel junctions. With the external magnetic field, the superconductivity in aluminum can be suppressed, thus changing the sample configuration from SIS' to NIS even at the lowest temperatures.

\begin{figure}[h]
\centering
\includegraphics[width=0.9\columnwidth]{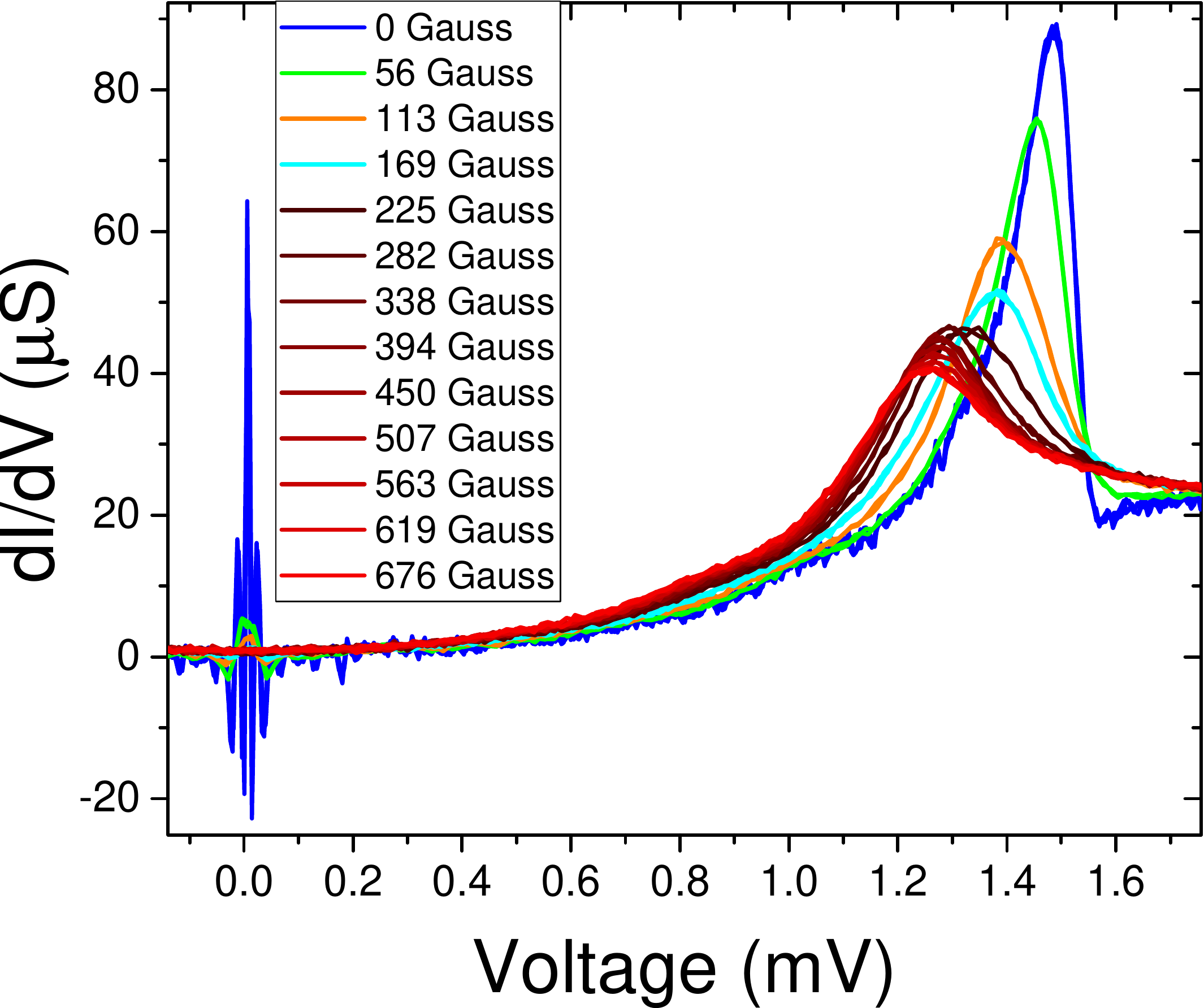}
\caption{Measured $dI/dV-V$ data at different magnetic fields at 140 mK temperature. Above $\sim 300$ G, when aluminum has turned normal, the Nb DOS  does not change appreciably.}
\label{nb_gap_mag_field}
\end{figure}

Fig. \ref{nb_gap_mag_field} shows the measured response of the junction, plotted as differential conductance $dI/dV$ vs bias voltage, with varying external magnetic field strength, at a refrigerator temperature of $T= 140$ mK. Concentrating first at the zero field data, one observes a clear supercurrent feature at zero bias, and another clear quasiparticle density of states (DOS) peak at around 1.5 mV. The position of the DOS peak is at the sum-gap $\Delta_{Nb}+\Delta_{Al}$ in the SIS' configuration. Clearly, the application of the field starts to suppress the supercurrent and shifts the DOS peak to a lower bias, as superconductivity in the Al electrode is gradually destroyed. At around 300 G, the shift of the peak stops, at which point Al has turned normal, and the peak position reflects only the size of the Nb gap $\Delta_{Nb} \sim 1.27$ meV. The amount of shift gives us then the value of the Al gap $\Delta_{Al} \sim 0.225$ meV. No extra Al gap feature at $V \sim \Delta_{Al}$ is seen, in contrast to our earlier devices where Nb was deposited first \cite{minna}.

Niobium would require a much higher field to suppress its superconductivity. This is also confirmed by the data, as at fields higher than 300 G very little change is seen in the $dI/dV$ curves of Fig. \ref{nb_gap_mag_field}. The conclusion is that the strengths of the external field used here do not directly influence superconductivity in the Nb lead. As the differential conductance is proportional to the quasiparticle DOS, no extra broadening or reduction of the gap due to pair-breaking caused by the magnetic field \cite{PhysRevLett.90.127001} is yet seen in Nb.  

\begin{figure*}[ht]
\includegraphics[width=1\textwidth]{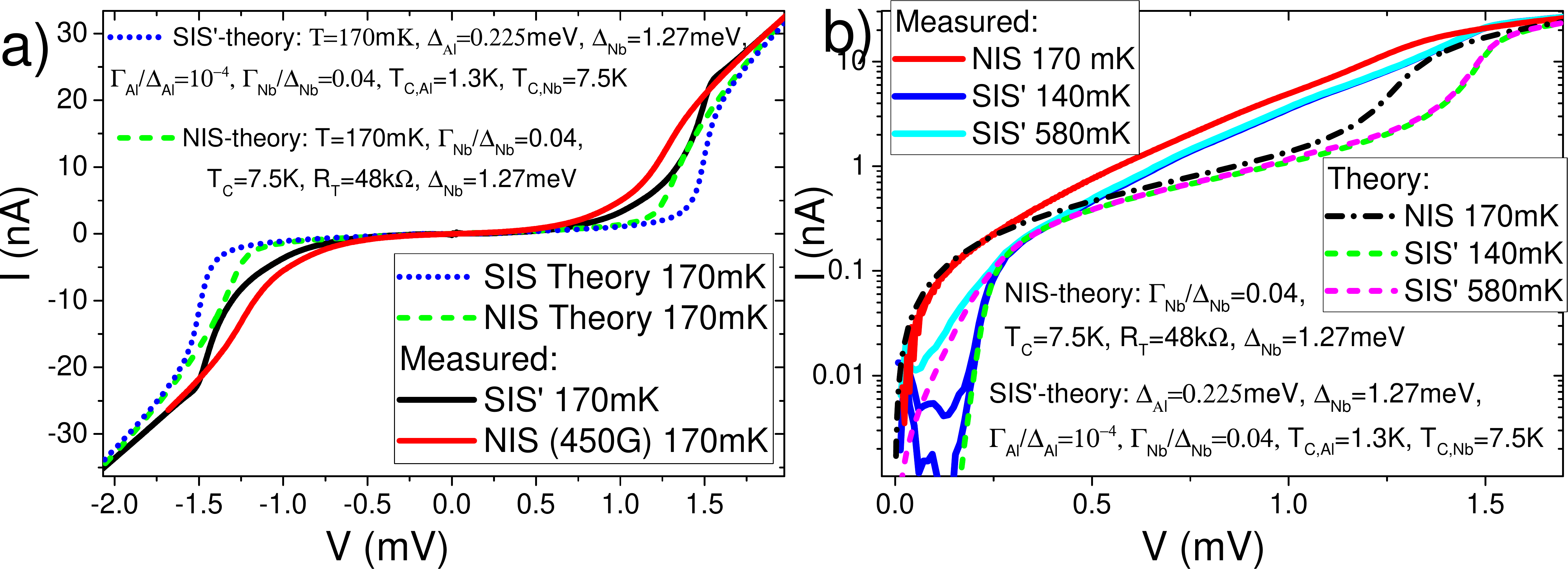}
\caption{((a) I-V measurements at $T=170$ mK with zero magnetic field (SIS') and 450 G (NIS) in linear scale. (b) I-V curves at log-linear scales, where SIS' data is shown at two different $T=140$ mK and  580 mK. Dotted lines show the simplest theoretical curves based on Eqs. \ref{NISeq}-\ref{SISeq} with the parameters shown.}
\label{nis_and_theory}
\end{figure*}
Next, in Fig. \ref{nis_and_theory} we compare the measured low-temperature $I-V$ characteristics  with theory both in the SIS' and NIS states, both in linear [Fig. \ref{nis_and_theory} (a)] and log-linear [Fig. \ref{nis_and_theory} (b)] scales. The measured $I$-$V$ curves show the expected strong non-linearities due to the gap(s), clearly seen in Fig. \ref{nis_and_theory} (a).

Theoretical plots are based on single-particle tunneling, which for the case of an NIS junction can be written in the symmetrized form \cite{SINIS} as
\begin{equation}
 I_{NIS}=\frac{1}{2eR_T}\int_{-\infty}^{\infty}\!\! d\epsilon N_{S}(\epsilon )[f_{N}(\epsilon-eV )-f_{N}(\epsilon+eV )],
\label{NISeq}
\end{equation}
where  $R_T$ is the tunneling resistance, $f_{N}(\epsilon)$ the Fermi function in the normal metal wire, and $N_{S}(\epsilon )$ is the normalized broadened  superconducting quasiparticle DOS  in the Dynes  model \cite{PhysRevLett.53.2437} 
\begin{equation}
 N_{S}(\epsilon, T_{S})  =\left | {\rm Re} \left ( \frac{\epsilon+i\Gamma}{\sqrt{(\epsilon+i\Gamma)^{2}-\Delta^{2}(T_S)}} \right ) \right |, 
\label{Dyneseq}
\end{equation}
with $\Gamma$ the broadening (Dynes) parameter and $\Delta(T_S)$ the temperature-dependent Nb gap. The above equation stresses the fact that in an NIS junction, current depends directly only on the temperature of the normal metal, not on the temperature of the superconductor (the temperature dependence of the gap does influence indirectly). 

For an SIS' junction, the current depends on the densities of states of both superconductors $N_{Nb}(\epsilon)$ and  $N_{Al}(\epsilon)$ and both Fermi-functions:


\begin{equation}
 I_{SIS'}\!\!=\!\!\! \frac{1}{eR_T} \!\! \!  \int_{-\infty}^{\infty}\!\!\!\!\!\!\!\!\!    d\epsilon N_{Nb}(\epsilon) N_{Al}(\!\epsilon+eV\!)(f_{Nb}(\epsilon)-\! f_{Al}(\!\epsilon+eV)),
\label{SISeq}
\end{equation}

where each superconductor has its own Dynes parameter $\Gamma_{Nb}$ and $\Gamma_{Al}$. 

The theoretical curves in Fig. \ref{nis_and_theory} were calculated with Eqs. \ref{NISeq} - \ref{SISeq}, using the known values of zero temperature gaps $\Delta_{Nb}$ and $\Delta_{Al}$ and the known value of the bath temperature (assumed to be equal to the normal metal and the superconductor temperature), and only varying the unknown Dynes parameters. The Dynes parameters are well constrained above by the data, as increasing $\Gamma$ will increase the current close to zero bias. 

We clearly observe in Fig. \ref{nis_and_theory} that the theoretical curves only fit the data close to the zero bias and above the gap edge at $V > 1.5$ mV, but cannot explain the data at an intermediate bias voltage range. A larger Dynes parameter for Nb cannot improve the fit, it would only lead to a stronger deviation, namely a too high current at bias close to zero and still a too low current at an intermediate range.  On the other hand, overheating, i.e. an effective temperature higher than the bath temperature is not realistic either. This can be seen by looking at the SIS' data in Fig. \ref{nis_and_theory} (b) at voltages below the step feature at 0.225 mV (Al gap), where the fits are actually good using the known bath temperatures. We thus believe that the excess current at around 0.5 - 1.5 mV has some other unknown origin. Note that the Al gap feature in the SIS' case is fully predicted by the single-particle tunneling theory, and evoking higher-order Andreev processes \cite{pashkin} is not required. In general, higher-order processes are very much suppressed in junctions with such high tunneling resistances as here ($R_T \sim 	50$ k$\Omega$).

\begin{figure*}[]
\includegraphics[width=1\textwidth]{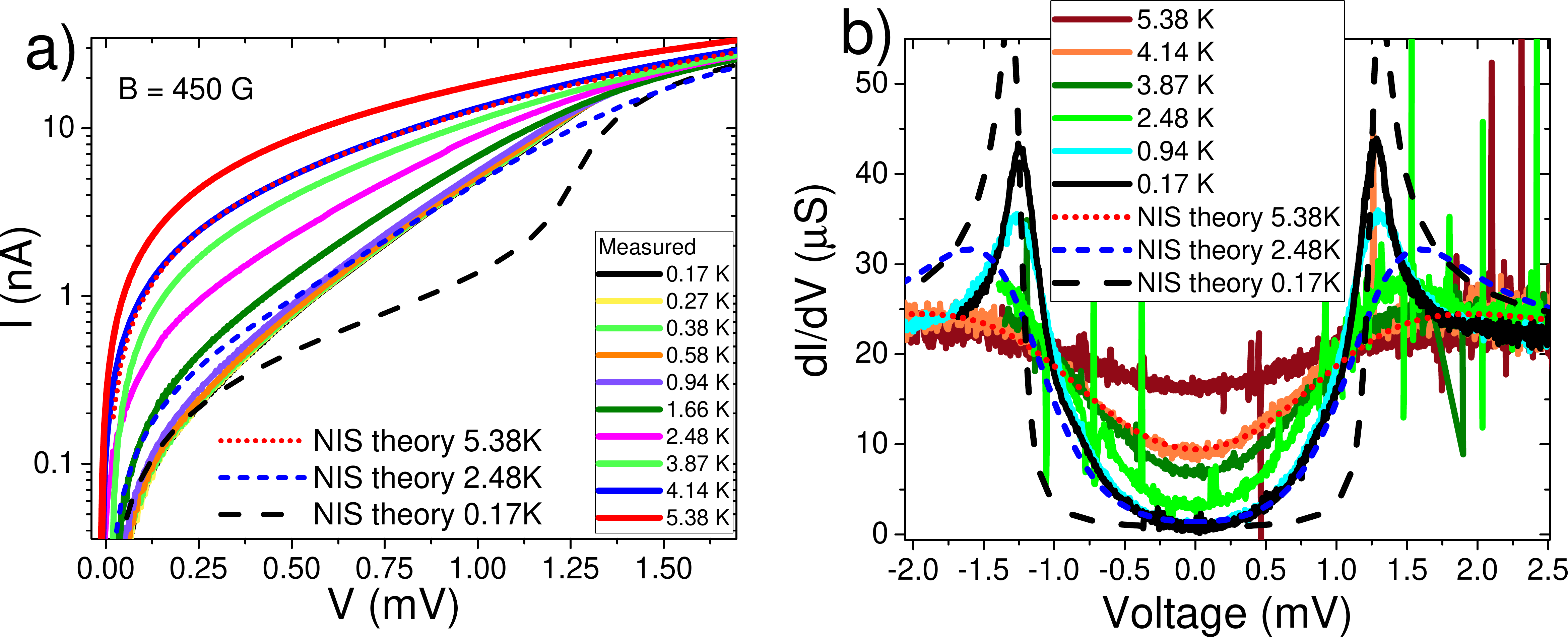}
\caption{(a) Measured temperature dependence of the (a) $I-V$ characteristics and (b) $dI/dV-V$ characteristics in the NIS state, when external field $B= 450$ G is applied.  Lines show theoretical results based on Eqs. \ref{NISeq}-\ref{Dyneseq}, with parameters $\Delta_{Nb} = 1.27$ meV, $\Gamma_{Nb}/\Delta_{Nb}= 0.04$ and temperatures $T=0.17$, 2.48 and 5.38 K. Note that apparent good fit of the 5.38 K theory is to the 4.14 K data.}
\label{NIS_T}
\end{figure*}

Let us now focus on the NIS data, where Al superconductivity is fully suppressed by the magnetic field. A set of measured $I-V$ characteristics with varying bath temperature, and some corresponding $dI/dV-V$  curves, are shown in Fig. \ref{NIS_T}. We observe that there is a clear temperature dependence, but only above $\sim 1$ K. It appears that the excess broadening saturates the response at temperatures below 1 K. The examples of theoretical curves shown in Fig. \ref{NIS_T} confirm that fits are not satisfactory at {\em any} temperature, there is always excess current in the experimental data, even at the highest temperatures. This is consistent with physical additional sub-gap states within the Nb gap.

\section{Thermometry applications}

\begin{figure}[]
\centering
\includegraphics[width=0.9\columnwidth]{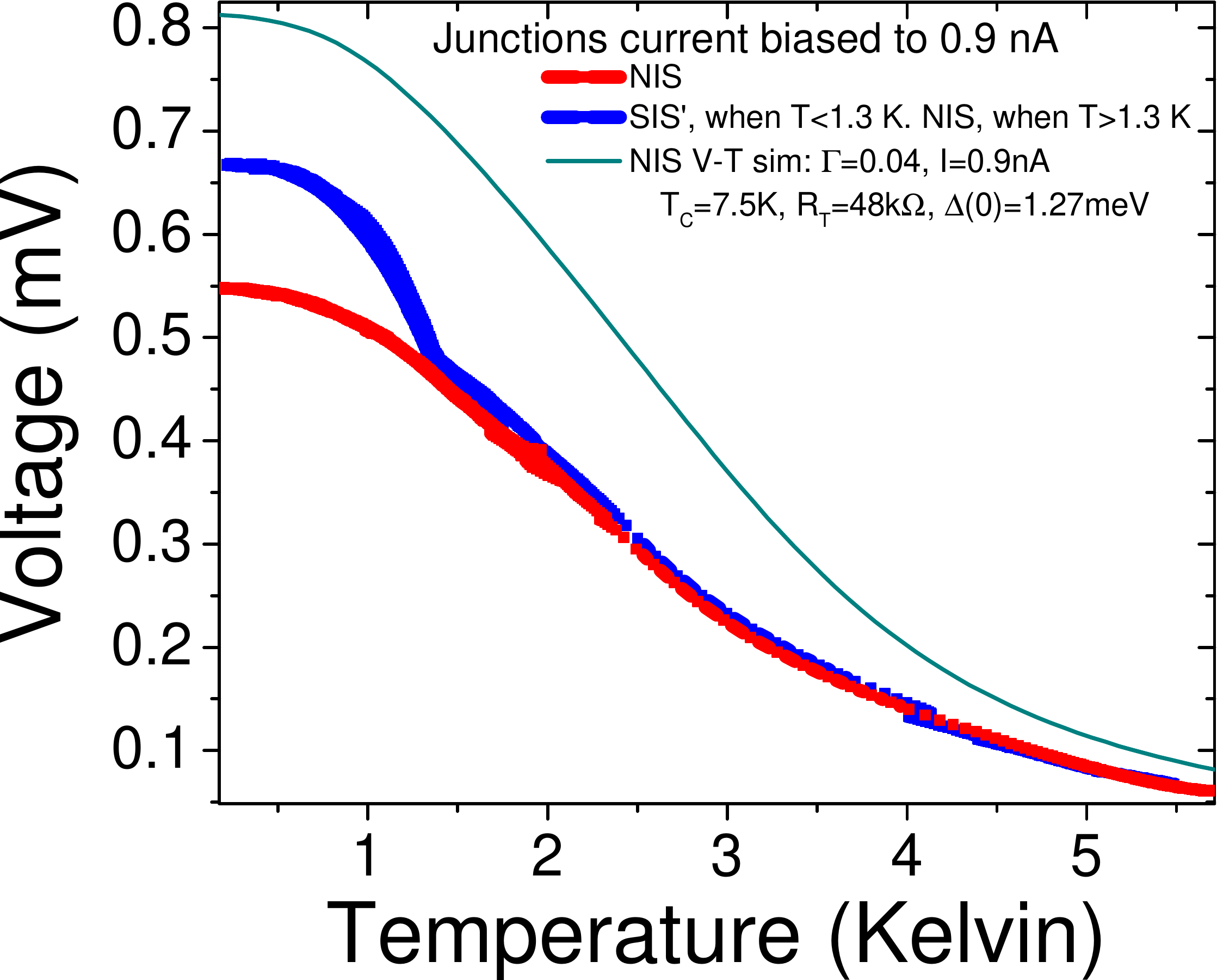}
\caption{Voltage vs. Temperature dependence of both NIS and SIS' junctions when constant current biased to 0.9 nA. A simple NIS  theory is shown as the solid line, with the same parameters as the $I$-$V$ theory. }
\label{VT_voltage_vs_temperature}
\end{figure}

Although the simplest theories could not be used to fit the electrical characteristics in the NIS state, there was still a clear temperature dependence observed. This means that sensitive nanoscale thermometry is still possible with these type of junctions. We also checked the thermometric responsivity, see Fig.\ref{VT_voltage_vs_temperature}, by constant current biasing the Al-AlOx-Nb device just below the Nb gap value at 0.9 nA,  and by measuring the bath temperature dependent voltage across the junction, as the refrigerator temperature was varied between 100 mK and 6.5 K. Fig.\ref{VT_voltage_vs_temperature} shows both NIS (finite $B$) and SIS' ($B=0$) data, in addition to the theory curve for the NIS case based on Eqs. \ref{NISeq}-\ref{Dyneseq}, with the same parameters that were used in the $I$-$V$ fits. As expected, the NIS and SIS' data are identical at temperatures above the aluminum $T_C = 1.3 $ K. The measured responsivity $dV/dT \sim 0.13$ mV/K is good and comparable to the responsivity of the Nb-AlOx-Cu NIS junctions\cite{minna}.  Moreover,  the responsivity is clearly enhanced by about a factor of two in the SIS' state at temperatures 1.0 K - 1.3 K. The simplest theory, as in the case of the $I-V$ curves, cannot describe the experiment well, except for the saturation caused by the Dynes broadening at $T < 1$ K.     

We conclude that Al-AlOx-Nb junctions fabricated in the way described here are actually quite promising for thermometry and bolometry \cite{schmidt,PhysRevApplied.3.014007} applications in the temperature range 1 K - 7.5 K. This is quite useful as (i) standard NIS thermometers with Al as the superconductor are limited to work below 1K, and (ii) previous NIS devices with Nb \cite{minna}, NbN \cite{nbn} and TaN \cite{tan} as the superconductor require much more difficult sample fabrication techniques, making  it much harder to integrate those devices with some other structures, for example  for thermal transport measurements \cite{Zen}. We believe Al-AlOx-Nb junctions are poised to do that in the near future, extending the temperature range of, for example, the phonon transport studies \cite{Zen} by almost an order of magnitude. On the other hand, NIS cooling with the observed broadening of the DOS is not currently feasible.     

\section{Discussion and Conclusions}

We have developed here a way to successfully fabricate sub-micron Al-AlOx-Nb tunnel junctions using standard e-beam resists and angle evaporation, with high quality Nb as determined by the measured values of the critical temperature $T_C \sim 7.5$ and niobium gap $\Delta \sim 1.3 $ meV, with the help of double oxidized tunneling barrier. The devices show great promise for local nanoscale thermometry in the temperature range 1 K - 7.5 K. 

The simplest theories for single particle tunneling did not, however, manage to describe the experimental data well, neither in in the NIS state where magnetic field was used to suppress superconductivity in Al, nor in the SIS' state. The experiment showed excess current, which could be interpreted as arising from additional quasiparticle states in the gap of Nb. We showed that these states cannot be described by the Dynes model \cite{PhysRevLett.53.2437}, which quite often can be used to describe the broadening of the density of states, either due to enivronmentally induced photon-assisted tunneling \cite{PhysRevLett.105.026803}, or due to quasiparticle lifetime effects \cite{PhysRevLett.53.2437}. We also showed that the pair-breaking by the applied magnetic field was not responsible for the broadening either, nor could higher-order (Andreev) tunneling \cite{PhysRevLett.100.207002} be responsible because of the high value of tunneling resistance in our samples. So we are currently left with an unsatisfying conclusion that full understanding is lacking.

Similar conclusions have been reached by other groups, either for somewhat similar Al-AlOx-Nb devices as here \cite{toppari} or for Al-AlOx-V devices \cite{shimada}. In both those cases, as in this study, the higher $T_C$ material (Nb or V) was evaporated on top of the tunneling barrier. In Ref. \cite{toppari}, a model of gap fluctuations within the tunneling barrier was developed to explain the additional broadening, whereas in Ref. \cite{shimada} detailed interpretation was not attempted. An important point we want to make in this regard here is that we {\em did not} see these extra states in the Nb-Al-AlOx-Cu samples \cite{minna}, where the lithography and Nb evaporation conditions (same evaporator) were practically identical. We thus have to conclude that material properties intrinsic to Nb, such as gap fluctuations, bulk impurity states etc. cannot explain the extra states in our experiment, and it may be quite relevant that in the experiments of Ref. \cite{minna}, Nb was evaporated first and is not in direct contact with the barrier. It is possible that non-idealities are generated in the barrier itself if Nb is deposited on top, or that the interface of Nb with AlOx forms interfacial states that are responsible for the gap states, for example by the mechanism of magnetic scattering (Shiba-theory) \cite{proslier}. Further studies on this issue are needed.

\ack

We acknowledge help in the experiments by A. Torgovkin, and help in the computational work by S. Chaudhuri and Z. Geng.  
This study was supported by the Academy of Finland project number 260880 and the Finnish Cultural Foundation.

\section*{References}
\bibliographystyle{unsrt}
\bibliography{cites}	
\end{document}